\documentclass[12pt,preprint]{aastex}

\shorttitle{Stability and Habitability of $\gamma$ Cephei}
\shortauthors{Haghighipour}

\begin{document}

\title{Dynamical Stability and Habitability of $\gamma$ Cephei \\
Binary-Planetary System}

\author{Nader Haghighipour}
\affil{Institute for Astronomy and NASA Astrobiology Institute,\\
University of Hawaii-Manoa, 2680 Woodlawn Drive,
Honolulu, HI 96822}

\email{nader@ifa.hawaii.edu}

\begin{abstract}

It has been suggested that the long-lived 
residual radial velocity variations observed in the
precision radial velocity measurements of the primary of 
$\gamma$ Cephei (HR8974, HD222404, HIP116727) 
are likely due to a Jupiter-like planet around this star \citep{Hatzes03}.
In this paper, the orbital dynamics of this plant is studied and also
the possibility of the existence of a hypothetical Earth-like planet 
in the habitable zone of its central star is discussed.
Simulations, which have been carried out for different values of the 
eccentricity and semimajor axis of the binary, as well as
the orbital inclination of its Jupiter-like planet, 
expand on previous studies of this system and
indicate that, for the values of the binary eccentricity smaller than 0.5,
and for all values of the orbital inclination of the Jupiter-like
planet ranging from 0$^\circ$ to 40$^\circ$, the orbit of this planet is 
stable. For larger values of the binary eccentricity, the system
becomes gradually unstable. Integrations also indicate that,
within this range of orbital parameters, a hypothetical Earth-like 
planet can have a long-term stable orbit only at distances 
of 0.3 to 0.8 AU from the primary star. The habitable zone of the
primary, at a range of approximately 3.1 to 3.8 AU, is, however, unstable. 
\end{abstract}


\section{Introduction}

Among the currently known extrasolar planet-hosting stars, 
approximately 20$\%$ are members of binaries or multistar
systems (Table 1)\footnote{See http://www.obspm.fr/planets for a complete 
and up-to-date list of extrasolar planets with their 
corresponding references.}. With the exception of
the pulsar-planetary system PSR B1620-26 \citep{sigurdsson03,richer03,beer04}, 
and possibly the newly discovered system HD202206 \citep{correia05}, 
the planets in these systems revolve only around one of the stars.
These systems are mostly 
wide with separations between 250 to 6500 AU.
At such large distances, the gravitational influence of the farther companion
on the dynamics of planets around the other star
is un-substantial. Simulations of the orbital
stability of a Jupiter-like planet around a star of
a binary system have shown that the existence of the
farther companion will have considerable effect if the separation
of the binary is less then 100 AU \citep{Jim02}. 
At the present, there are three planet-hosting binary/multistar systems
with such a separation; $\gamma$ Cephei \citep{Hatzes03}, 
GJ 86 \citep{Els01}, and HD188753 \citep{Konacki05}.
This paper focuses on the dynamics, long-term stability, and the
habitability of $\gamma$ Cephei.

To many observers, the discovery of a planet in a dual-star system 
is of no surprise. There are many observational evidence that indicate
the most common outcome of the star formation
process is a binary system \citep{Math94,White01}. There is also
substantial evidence for the existence of potential planet-forming
circumstellar disks in multiple star systems 
\citep{Math94,Akeson98,Rodriguez98,White99,Silbert00,Math00}. 
To dynamicists, on the other hand, the discovery of a planet in a binary
star system marks the beginning of a new era of more challenging questions.
Three decades ago, models of planet formation in binary 
systems did not permit planet growth in binaries with separation comparable
to those of $\gamma$ Cephei, GJ 86, and HD188753 \citep{Hep74,Hep78}.
Results of recent simulations by \citet{Nelson00} also agree with
those studies and imply that planets cannot grow via either core-accretion
or disk instability mechanisms in binaries with separation of approximately 50 AU.
Recent discoveries of planets in dual star systems, however,
have cast doubt in the validity of those theories, and have now confronted 
astrodynamicists with new challenges. Questions such as,
how planets are formed in binary star systems, 
what are the criteria for their long-term stability, 
can such planet-harboring systems be habitable, 
and how are habitable planets formed in binary star systems, are 
no longer within the context of hypothetical systems and
have now found real applications.

Theorizing the formation of a planet in a dual star system
requires a detailed analysis of planet formation
at the presence of a stellar companion.
Such a study is beyond the scope of this paper. However,
for the sake of completeness, papers by 
\citet{Boss98,Boss04, Boss05}, \citet{nelson03}, and \citet{mayer04}
on the effect of a stellar companion on the dynamics of a 
planet-forming nebula, and articles by 
\citet{Marzari00}, \citet{Barbieri02}, \citet{Quintana02}, and
\citet{Lissauer04} 
on the formation of Jupiter-like and 
terrestrial planets in and around binary star systems are cited.

Prior to constructing a theory for the formation of planets
in binary star systems, it proves useful to develop a
detailed understanding of the dynamics of planets in such environments. 
This is a topic that
despite the lack of observational evidence, has
always been of particular interest to dynamicists. For instance,
in 1977, in search of criteria for the stability of planets
in binary star systems, Harrington carried out a study of
the orbital stability of Jupiter- and Earth-like planets
around the components of a binary.
In his simulations, Harrington considered two equal-mass stars 
and numerically integrated the equations of motion of a planet on a 
circular orbit in and around a binary with an eccentricity
of 0, and 0.5. As expected, Harrington's results indicated 
that planets can, indeed, have stable orbits in binary star
systems provided they are either sufficiently close to their host
stars, or sufficiently far from the entire binary system
\citep{Har77}. 

Stability of planets in binary star systems has also
been studied by \citet{Black81}, \citet{Black82}, and \citet{Black83}.
In an effort to establish criteria
for the orbital stability of three-body systems,
these authors studied the survival time of a planet in the gravitational
field of two massive bodies and mapped the parameter-space 
of the system for planetary orbits around each of 
the stars, as well as the entire binary system.
Their results indicated that, in binary systems
where the stellar components have comparable masses, orbital inclination
of the planet will not have significant effect on its stability.
A result that had also been reported by \citet{Har77}.
However, when the mass of one of the components of the binary is comparable
to the mass of Jupiter, planetary orbits with inclinations higher than 
50$^\circ$ tend to be unstable.

To study the orbital stability of planetary bodies,
different authors have used different stability criteria. For instance, 
the notion of stability as introduced by \citet{Har77}
implied no secular changes in the semimajor axis and orbital
eccentricity of a planet during the time of integration. 
\citet{Szeb80}, and \citet{Szeb81}, on the other hand,
used the integrals of motion and curves of zero velocity to
establish orbital stability. These authors considered
a restricted, planar and circular three-body 
system with a small planet (with negligible mass) orbiting
either of the stars, or the entire binary system.
Allowing arbitrary perturbations in the equation of
motion of the planet, they mapped the parameter-space of the system 
(i.e., orbital radius, vs. ratio of the mass of the smaller 
component to the total mass of the binary) and identified
regions where the orbit of the planet could be Hill stable. 
In the present paper, the orbital
eccentricity of an object and its distance to other bodies 
of the system are used to set the criteria for stability.
The orbit of an object is considered stable if, 
for the entire duration of integration, it's orbital eccentricity
stays below unity, it doesn't collide with other bodies,
and it doesn't leave the gravitational field of the system.

Orbits of planets in binary star systems can be divided into 
different categories. In an article in 1980, Szebehely 
distinguished these categories as: Inner orbits,
where a planet revolves around the primary star, satellite orbits,
where a planet revolves around the secondary star, and outer orbits,
where a planet revolves around the entire binary system \citep{Szeb80}.
Another classification 
has also been reported by \citet{Dvorak83}. As noted by this author, 
the systematic study of
the stability of resonant periodic orbits 
in a restricted, circular, three-body system 
by \citet{Henon70} implies three
types of planetary orbits in a binary system; the S-type
where the planet revolves around one of the stars, the P-type
where the planet orbits the entire binary systems, and the
L-type where the planet has a stable librating orbit around
$L_4$ and $L_5$ Lagrangian points. 
According to this classification, the dual star system of
$\gamma$ Cephei is an S-type binary-planetary system. 

Extensive studies have been done on the dynamical stability
of S-type binary-planetary systems
\citep{Dvorak88a,Benest88,Benest89,Benest93,Benest96,Holman97,Holman99,
Pilat02,Dvorak03,Dvorak04,Pilat04,Musielak05}. 
Although in these articles, the stability of S-type
systems has been studied for different values of the binary's
mass-ratio and orbital parameters, simulations
have been limited to restricted
cases such as co-planar orbits, similar-mass binary components, and
circular planetary orbits, and/or the durations of simulations have been
no more than tens of thousands of the binary's orbital period.
A more detailed analysis of the stability of binary-planetary systems,
particularly within the context of habitability, however, requires 
simulating the orbital dynamics of these systems for longer times.
This paper extends previous studies by focusing on
(1) the study of the long-term stability of $\gamma$ Cephei 
binary-planetary system, and (2) identifying regions of its parameter-space
where, in the habitable zone of its primary star and
at the presence of its Jupiter-like planet, an Earth-like object
can have a long-term stable orbit.
In this study, simulations are extended to a larger parameter-space
where the orbital elements of the binary and the inclination
of the planets' orbits are included, 
and the stability of the system is studied for ten to
hundred million years.

The outline of this paper is as follows.
In $\S$ 2, the initial set up for the numerical integration of the system
is presented. The results of the numerical simulations
are given in $\S$ 3, and in 
$\S$ 4  the habitability of the system is discussed.
Section 5 concludes this study by reviewing the results and
comparing them with previous studies.

\section {Initial Set Up}

The dual-star system of $\gamma$ Cephei is a spectroscopic binary 
with a 1.59 solar-mass  K1 IV subgiant as its primary \citep{Fuhr03}
and a probable red M dwarf, with a mass-range  of 0.34 to 0.78 solar-mass
\citep{Endl05}, 
as its secondary. The semimajor axis and eccentricity of this system
are, respectively, $18.5 \pm 1.1$ AU and $0.361 \pm 0.023$, as reported by
\citet{Hatzes03}, and $20.3 \pm 0.7$ AU and $0.389 \pm 0.017$,
as reported by \citet{Griffin02}. The primary
star of this system has been suggested to be the host 
to a planet with a minimum mass of 1.7 Jupiter-mass, on an 
orbit with semimajor axis of $2.13 \pm 0.05$ AU, and eccentricity
of $0.12 \pm 0.05$ \citep{Hatzes03}. 

The existence of two sets of reported values for the
orbital semimajor axis and eccentricity of this binary, and also
a mass-range for its secondary component, 
have caused $\gamma$ Cephei to have a large parameter-space. 
This parameter-space that consists of
the binary's semimajor axis $(a_b)$ and eccentricity $(e_b)$, 
the planet's orbital inclination with respect to the plane 
of the binary $(i_p)$, and the binary's
mass-ratio $\mu={m_2}/({m_1}+{m_2})$ 
with $m_1$ and $m_2$ being the masses of the primary
and secondary stars, respectively, is the space of the
initial conditions for numerical integrations of the system. 
The first goal of this
study is to identify regions of this parameter-space where the
Jupiter-like planet of the system can have long-term stable orbits.

An important quantity in determining the stability of a planet in
a binary star system is the planet's semimajor axis. \citet{Dvorak88a}, and 
\citet{Holman99} have obtained an empirical formula for
the maximum value of the semimajor axis of a stable planetary
orbit ({\it critical} semimajor axis, $a_c$)
 in terms of the binary mass-ratio and orbital eccentricity
in a co-planar S-type binary-planetary system. As shown by these authors,
\begin{eqnarray}
&{{a_c}/{a_b}}=(0.464\pm 0.006)+
(-0.380 \pm 0.010)\mu + (-0.631\pm0.034) {e_b}\nonumber\\
&\qquad\qquad\qquad
+ (0.586 \pm 0.061) \mu {e_b} + (0.150 \pm 0.041) {e_b^2} 
+(-0.198 \pm 0.047)\mu {e_b^2}\>.
\end{eqnarray}
\noindent
Since, due to their uncertainties, 
the reported values of $a_b$, $e_b$, and $\mu$ 
for $\gamma$ Cephei binary system vary within
certain ranges, the value of the planet's critical semimajor
axis will also vary within a range of values.
Figure 1 shows the graph of $a_c$ in terms of the binary 
eccentricity for different reported values of $a_b$
(including its corresponding uncertainties), and
also for all permutations of $\pm$ sings of Eq. (1).
The value of $\mu$ in this graph is 0.2 corresponding
to a mass of 0.4 solar-mass for the farther companion.
Figure 1 shows that, in a co-planar system,
for any given value of the binary eccentricity,
the orbit of the Jupiter-like planet will be stable as long as
the value of its semimajor axis stays below the minimum
value of its corresponding range of $a_c$. In fact,
the lower boundary of this graph makes the 
upper limit of the admissible values of the planet's
semimajor axis for which the planet's orbit will be stable.

Although Fig. 1 presents a general idea of the stability of the
$\gamma$ Cephei's planet for low or zero orbital inclination, 
in order to portray a more detailed 
picture of the dynamical state of this body, and for the
purpose of extending the analysis to more general cases which
include inclined orbits as well,
and also, to better understand the dynamical effects
of this planet on the long-term stability of a habitable planet
in this system, numerical simulations were carried out for different 
values of the semimajor axis and eccentricity of the
binary, as well as the orbital inclination of the planet. The
initial value of $e_b$ was chosen from the range of 0.2 to 0.65
in increments of 0.05, and the initial orbital inclination of 
the planet was chosen from the values of 
${i_p}=$0, 2$^\circ$, 5$^\circ$, 10$^\circ$, 20$^\circ$, 40$^\circ$, 
60$^\circ$, and 80$^\circ$. 
Numerical simulations were carried out for different values of
${a_b}$ ranging from 18 to 22 AU.

\section {Stability of the Jupiter-like Planet}

The three-body system of $\gamma$ Cephei binary-planetary system
was integrated numerically using a conventional Bulirsch-Stoer integrator.
Integrations were carried out for different values of
${a_b}, \, {e_b}$ and $i_p$, as indicated in the previous section.
Table 1 shows the initial orbital parameters of the
system. For the future purpose of integrating the equation of
motion of an Earth-like planet within a large range of distances
from the primary star, the timestep of each simulation was set
to 1.88 days, equal to 1/20 of the orbital period of a 
planet at a distance of 0.3 AU from the primary star. This timestep
was used in all orbital integrations of the system.

Figure 2 shows the results of integrations for a co-planar system 
with $\mu=0.2$ and for different values of the binary  eccentricity.
The initial value of the semimajor axis of the binary 
is 21.5 AU. As shown here, the system is stable for 
$0.2 \leq {e_b} \leq 0.45$.
This result is consistent with
the stability condition depicted by Fig. 1. 
Integrations also indicate that
the system becomes unstable in less than a few thousand years
when the initial value of the binary eccentricity exceeds 0.5. 

To investigate the effect of planet's orbital inclination $({i_p})$
on its stability, the system was also 
integrated for different values of $i_p$.
The results indicate that for $0.20 \leq {e_b} \leq 0.45$,
the system is stable for all values of planet's orbital inclination
less than 40$^\circ$.  
Figure 3 shows the semimajor axes and orbital eccentricities
of the system for ${e_b}=0.2$ and for  $i_p$=5$^\circ$, 10$^\circ$, and 20$^\circ$. 
For orbital inclinations larger than 40$^\circ$, the
system becomes unstable in a few thousand years.

\subsection {Kozai Resonance}

An exception to the above-mentioned instability condition
was observed for the planet's inclination equal to 60$^\circ$.
At this orbital inclination and
for the initial eccentricities of 0.25 and 0.17
for the binary and planet, respectively, 
the system showed the signs of a Kozai resonance. Figure 4 shows the
semimajor axes and eccentricities of the binary and planet in this case.

As demonstrated by \citet{Kozai62}, the exchange 
of angular momentum between the small
body of the system (here, the Jupiter-like planet) and the binary,
can cause the orbital eccentricity of the planet to reach high values at 
large inclinations (Fig. 4). Averaging the equations of motion
of the system over mean anomalies, one can show that in this case,
the averaged system is integrable when the ratio of distances are 
large \citep[the Hill's approximation,][]{Kozai62}.
The Lagrange equations of motion in this case, indicate that,
to the first order of planet's eccentricity, the longitude of
the periastron of the planet, $\omega_p$, librates
around a fix value. Figure 4 also shows $\omega_p$ for the 
duration of integration. As shown here, this quantity librates 
around 90$^\circ$.  

In a Kozai resonance, the longitude of periastron and the 
orbital eccentricity of the small body are related to its 
orbital inclination as \citep{Innanen97}
$$
{\sin^2}{\omega_p}=\,0.4\,{\csc^2}{i_p},
\eqno (2)
$$
\noindent
and
$$
{(e_p^2)_{\rm max}}={1\over 6}\,\Bigl[1-5\cos (2{i_p})\Bigr].
\eqno(3)
$$
\noindent
From Eq. (2), one can show that the Kozai resonance may occur
if the orbital inclination of the small body is
larger than 39.23$^\circ$. As mentioned above, in this study, 
the Kozai resonance occurred for ${i_p}={60^\circ}$.
For the minimum value of ${i_p}$, the maximum value
of the planet's orbital eccentricity, as given by Eq. (3), 
is equal to 0.764. Figure (4) also shows that
$e_p$ stays below this limiting value at all times.

As shown by \citet{Kozai62} and \citet{Innanen97}, 
in a Kozai resonance, the disturbing function
of the system, averaged over the mean anomalies, is independent of the
longitudes of ascending nodes of the small object and perturbing
body. As a result, the quantity ${\sqrt{{a}(1-{e^2})}}\,\cos {i}$
(shown as the ``Reduced Delaunay Momentum'' in Fig. 4)
becomes a constant of motion. Figure 4 shows this quantity
for the Jupiter-like planet of the $\gamma$ Cephei system. 
Since the eccentricity and inclination of the planet vary with time, the
fact that the quantity above is a constant of motion implies that the
time-variations of these two quantities have the same periods
and they vary
in such a way that when $i_p$ reaches its maximum, $e_p$ reaches
its minimum and vice versa. Figure 5 shows this clearly.

\section {Habitability}

To study the habitability of $\gamma$ Cephei, one has to
investigate the long-term stability of a habitable planet in
the habitable zone of this system. 
A habitable zone is commonly referred to a region around a star 
where an Earth-like planet can maintain liquid water on its surface.
The capability of maintaining liquid water depends on several factors
such as
the amount of radiation that such a planet receives 
from the star. This radiation itself depends on the star's luminosity,
and vary with its radius $R$ and surface temperature $T$ as,
$$
F(r)={1\over {4\pi}}L(R,T){r^{-2}}=\sigma{T^4}{R^2}{r^{-2}}.
\eqno (4)
$$
\vskip 3pt
\noindent
In this equation, $L(R,T)$ is the luminosity of the star,
$\sigma$ is the Boltzmann constant, and  $F(r)$ is the apparent
brightness of the star denoting the
amount of stellar radiation that, in a 
unit of time, is distributed over the unit area of a sphere with radius $r$.

Equation (4) indicates that the width of a habitable
zone and the locations of its inner and outer boundaries
vary with the physical properties of the star.
The inner edge of a habitable zone is defined as
the largest distance from a star where, due to 
photodissociation and runaway greenhouse
effect, the planet loses all its water.
The outer edge of a habitable zone, on the other hand, 
corresponds to the shortest distance from a star where water can no
longer exist in liquid phase and begins to freeze. 
In other words, the outer edge of a habitable 
zone corresponds to the largest distance from a star where an
Earth-like planet with a carbon-dioxide atmosphere 
can still, in average, maintain a temperature of 273 K on its surface
\citep{Kasting93}. As noted by \citet{Jones05}, such a definition
for the boundaries of a habitable zone is somewhat conservative, and
the outer edge of this zone may, in fact, be farther away
\citep{Forget97,Williams97,Mischna00}.
 
Since the above-mentioned definition of a  habitable zone
is based on the notion of habitability and life on  Earth,
one can use Eq. (4) to determine habitable regions around other stars
by comparing their luminosities with that of the Sun. 
That is, a habitable zone can be defined as a region around a star where  
an Earth-like planet can receive the same amount of radiation as Earth
receives from the Sun, so that it can develop and maintain similar habitable 
conditions as those on Earth. From Eq. (4), the statement above implies
\vskip 1pt
$$
{{F(r)}}\,=\,
{\Bigl({T\over {T_{\rm Sun}}}\Bigr)^4}\,
{\Bigl({R\over {R_{\rm Sun}}}\Bigr)^2}\,
{\Bigl({r\over{r_{\rm Earth}}}\Bigr)^{-2}}\,
{{F_{\rm Sun}}}({r_{\rm Earth}})
\eqno (5)
$$
\vskip 3pt
\noindent
where now $F(r)$ represents the apparent brightness of a star
with a luminosity of $L(R,T)$ as observed from an Earth-like
planet at a distance $r$ from the star, $r_{\rm Earth}$
is the distance of Earth from the Sun, and ${F_{\rm Sun}}({r_{\rm Earth}})$
represents the brightness of the Sun at the location of Earth.

The primary star of $\gamma$ Cephei binary system has a temperature
of 4900 K and a radius of 4.66 solar-radii \citep{Hatzes03}. Considering 
5900 K as the surface temperature of the Sun,
Eq. (5) indicates that in order for Earth to receive the same amount
of radiation as it receives from the Sun, it has to be at a distance
of $r\sim$ 3.2 AU from the $\gamma$ Cephei's primary star.
The habitable zone of the Sun, on the other hand, is considered
to be between 0.95 to 1.37 AU \citep{Kasting93}. This range corresponds 
to a range of apparent solar brightness between 1.1 and 0.53, 
and implies that a similar region around the primary of $\gamma$ Cephei 
extends from 3.13  to 3.76 AU from this star.

The width and distance of the habitable zone of $\gamma$ Cephei 
have also been reported in papers by \citet{Dvorak03}, and \citet{Jones05}.
In a study of the stability of planets in the $\gamma$ Cephei binary
system, \citet{Dvorak03} have considered a range of 1 to 2.2 AU as
the habitable zone of the system's primary star. From Eq. (5), at these distances, 
the apparent brightness of this star varies between 10.3
to 2.1 times the brightness of the Sun at 1 AU.  
With such brightness, it is unlikely that an Earth-like
planet around the primary of $\gamma$ Cephei, at the range of distances 
reported by these authors, can maintain
similar habitable conditions as those on Earth.
More recently, \citet{Jones05} 
studied the habitability of extrasolar planetary systems
and tabulated the 
habitable regions of a large number of planet-hosting stars.
According to these authors, the habitable zone of
the primary of $\gamma$ Cephei extends from 2.07 to 4.17 AU from
this star \citep{Jones05}. These authors consider 
this range to be conservative and
mention that the outer boundary of the actual habitable zone may be somewhat larger. 
In the present study, however, the habitable zone of the primary of
$\gamma$ Cephei is considered to be narrower and 
between 3.1 to 3.8 AU from this star. 

Numerical integrations were carried out to study the stability
of an Earth-like planet in $\gamma$ Cephei system. Although the
habitable zone of the system was considered to be from 3.1 to 3.8 AU, 
stability of an Earth-like planet was studied
at different locations, ranging from 0.3 to 4.0 AU
from the primary star. Table 3 shows the ranges of the
orbital parameters of this object, as well as those of the 
binary and its Jupiter-like planet.
As shown in this table, numerical simulations were also
carried out for different values of the orbital inclinations
of Earth- and Jupiter-like planets.
For each arrangement of these bodies, the equations of motion
of a full four-body system consisting of the binary, its Jupiter-like 
planet, and an Earth-like object were numerically integrated.
Figure 6 shows the survival times of Earth-like planets
in terms of their initial positions for a co-planar 
arrangement with ${e_b}=0.3$.  
As shown here, an Earth-like planet will
not be able to sustain a stable orbit in the habitable zone of the
primary star. Results of numerical simulations 
indicate that the orbit of an Earth-like planet is stable when
$0.3 \leq {a_E} \leq 0.8$ AU, ${0^\circ} \leq {i_E}=
{i_p} \leq {10^\circ}$, and ${e_b}\leq 0.4$, where $a_E$ and $i_E$
represent the semimajor axis and orbital inclination of the
Earth-like planet, respectively.
Figures 7 shows the time-variations of the semimajor axes, eccentricities, and 
orbital inclinations of one of such four-body systems  
for ${i_p}=5^\circ$, and for 100 million years.

\section {Conclusion}

The results of a study of the orbital stability of the binary-planetary 
system of $\gamma$ Cephei have been presented. 
Numerical integrations of the full three-body system
of the binary and its Jupiter-like planet indicate 
that the orbit of this planet is stable for the values of the
binary eccentricity less than 0.5
[see Fig. 5 of \citet{Musielak05} for simulations of the system
for $e_b$ close to zero] and for the 
planet's orbital inclination up to 40$^\circ$. For larger values
of the inclination, the system becomes 
unstable except at 60$^\circ$ where the planet 
may be in a Kozai resonance. 

The focus of the first part of this study was
on the effects of the variations of the binary eccentricity
and planet's orbital inclination on the stability
of the system. For that reason, numerical simulations were
carried out for only one value of the semimajor axis
of the Jupiter-like planet. However, simulations by \citet{Pilat02}, 
\citet{Dvorak03}, and \citet{Pilat04} have indicated that the region of the stability of 
this planet extends to larger distances beyond its current location. 
\citet{Dvorak03} studied the stability of this planet by
numerically integrating the equation of motion of a massless
object in a restricted elliptical three-body system and showed that
its stable region extends to 4 AU from the primary star.
\citet{Pilat04}, on the other hand, have shown that, 
allowing a range of 0.1 to 0.9 for the binary's mass-ratio $(\mu)$ 
will limit this stable reigon to only 3.6 AU from the primary star.

In addition to the dynamics of its Jupiter-like planet,
the binary-planetary system of $\gamma$ Cephei was also
studied as a possible system for harboring habitable planets.
The habitability of this system
was studied by including a hypothetical Earth-like planet
at different locations from its primary star 
and integrating the equations of motion of a 
complete elliptical four-body system. Simulations
indicated that an Earth-like planet, initially on a circular
orbit, would have an
unstable orbit for the values of its semimajor axis larger
than 0.8 AU. The habitable zone of the system, between 3.1 to 3.8 AU,
is within this unstable region.

A report of the instability of an Earth-like planet for ${a_E}>0.8$ AU 
can also be found in the work of
\citet{Dvorak04}. In a study of the stability of a
fictitious massless planet in the vicinity of the Jupiter-like
planet of $\gamma$ Cephei, these authors extended two previous
studies by \citet{Dvorak03} and \citet{Pilat04}, and
showed that such an
object, when initially on a circular orbit, cannot maintain its 
stability between 1.7 to 2.6 AU from the primary star.
Considering an Earth-like planet as a massless object, and
simulating its dynamics in the $\gamma$
Cephei system, \citet{Dvorak03} and \citet{Pilat04} had already
shown that, when the inclination of
the Jupiter-like planet of the system varies from 
0$^\circ$ to 40$^\circ$, such an Earth-like object could have a 
stable orbit only in a region between 0.6 to 0.8 AU from the primary star. 
Their results had also indicated 
an island of stability at 1 AU from this star.

It is important to emphasize that, despite of some similarities 
between the results presented in this paper and those of
\citet{Dvorak03} and \citet{Pilat04}, the latter two studies do not
fully represent the dynamical state of the $\gamma$ Caphei system.
Studies of the stability of this system, 
as presented by these authors,
are limited to elliptic restricted three-body
cases. Integrations of the equation of motion of 
an Earth-like planet, in those articles, were also carried out
within the context of the motion of a massless particle in
an elliptic restricted system. 
These restrictions limit the generalization
of the results obtained by these authors, particularly, when
studying the habitability of the system.
In the present paper, however, these limitations were overcome 
by carrying out simulations for a complete elliptical four-body 
system. The results of these simulations indicated that
the range of the orbital stability of an Earth-like planet does not extend 
beyond 0.8 AU from the primary star, and 
the island of stability at 1 AU,
as reported by  \citet{Dvorak03} and \citet{Pilat04} is
indeed unstable. Simulations also indicated that, although
the region of the stability of an Earth-like planet
as reported by these authors (0.6 to 0.8 AU) is within  
the region of stability indicated in this paper (0.3 to 0.8 AU), 
unlike what they report, 
the orbital inclination of the Jupiter-like planet 
of the system cannot exceed 10$^\circ$. 

In closing, it is necessary to mention that
the study of the habitability of the primary star of $\gamma$ Cephei,
as presented in this paper, has one limitation. The evolution of this star
during the time of integration has not been 
taken into consideration. This star is a 3 billion years old K1 IV subgiant 
that is still in the process of approaching the giants' region
of the HR diagram. While this star expands, its luminosity increases,
and as a result, its habitable zone will move toward larger distances.
Although the results of numerical simulations
indicate that an Earth-like planet will have an unstable orbit 
at distances beyond the current habitable zone of the primary of
$\gamma$ Cephei, it is necessary to extend these
simulations to even larger distances, particularly when the
stability of Earth-like planets is studied for several hundred million years.
Additionally, it is important to investigate how different values of the mass
of the farther companion would affect the
dynamical stability and habitability of the system.
The results of such simulations are currently in preparation for publication.

\acknowledgments

I am thankful to the Department of Terrestrial Magnetism at the
Carnegie Institution of Washington for access
to their computational facilities where the numerical simulations
of this work were performed.
This work has been supported by the NASA Astrobiology
Institute under Cooperative Agreement NNA04CC08A at the Institute 
for Astronomy at the University of Hawaii-Manoa.

\clearpage
\begin{deluxetable}{llll} 
\tablewidth{0pt} 
\tablecaption{Extrasolar Planet-Hosting Stars in Binary Systems } 
\tablehead{ 
\colhead{\hskip -90pt Star}&
\colhead{\hskip -75pt Star}&
\colhead{\hskip -75pt Star}&
\colhead{\hskip -90pt Star}}
\startdata 
HD142    (GJ 9002)                & 
HD3651                            & 
HD9826   ($\upsilon$ And)         &    
HD13445  (GJ 86)                  \\
HD19994                           & 
HD22049  ($\epsilon$ Eri)         & 
HD27442                           & 
HD40979                           \\
HD41004                           & 
HD75732  (55 Cnc)                 & 
HD80606                           & 
HD89744                           \\
HD114762                          & 
HD117176 (70 Vir)                 & 
HD120136 ($\tau$ Boo)             &  
HD121504                          \\
HD137759                          & 
HD143761 ($\rho$ Crb)             & 
HD178911                          & 
HD186472 (16 Cyg)                 \\
HD190360 (GJ 777A)                & 
HD192263                          & 
HD195019                          & 
HD213240                          \\
HD217107                          & 
HD219449                          & 
HD219542                          & 
HD222404 ($\gamma$ Cephei)        \\
HD178911                          &
HD202206                          &
HD188753                          \\ 
PSR B1257-20                      &
PSR B1620-26                      &
\enddata 
\end{deluxetable}

\clearpage
\begin{deluxetable}{lcc} 
\tablewidth{0pt} 
\tablecaption{Initial Orbital Parameters} 
\tablehead{ 
\colhead{Parameters}  
&\colhead{Planet}
&\colhead{Binary}}
\startdata 
$a$(AU)                  & 2.13    & 18-22           \\
$e$                      & 0.17    & 0.20-0.65           \\
$i$(deg)                 & 0-80       & 0               \\
$\Omega$(deg)            & 0       & 0               \\
$\omega$(deg)            & 74      & 160              \\
$M$(deg)                 & 104     & 353              \\
\enddata 
\end{deluxetable}

\clearpage
\begin{deluxetable}{lccc} 
\tablewidth{0pt} 
\tablecaption{Initial Orbital Parameters of an Earth-like Planet} 
\tablehead{ 
\colhead{Parameters}
&\colhead{Earth-like Planet}  
&\colhead{Jupiter-like Planet}
&\colhead{Binary ($\mu=0.2$)}}
\startdata 
$a$(AU)               & 0.3-4.0    & 2.13          & 18,19,20     \\
$e$                   & 0        & 0.17          & 0.2,0.3,0.4  \\
$i$(deg)              & 0,2,5,10 & 2,5,10,20,40  & 0            \\
$\Omega$(deg)         &0         & 0             & 0            \\
$\omega$(deg)         &0         & 74            & 160          \\
$M$(deg)              &0         & 104           & 353          \\
\enddata 
\end{deluxetable}

\end{document}